\documentclass[traditabstract]{aa}
\usepackage{graphicx}
\usepackage{txfonts}
\usepackage{natbib}
\bibpunct{(}{)}{;}{a}{}{,} % to follow the A&A style
\newcommand{\src}{BR\,1202-0725}
\newcommand{\Msun}{M$_{\odot}$}
\newcommand{\Msunyr}{M$_{\odot}$\,yr$^{-1}$}
\newcommand{\La}{Ly$\alpha$}

\newcommand{\Htwo}{H$_2$}
\newcommand{\Cplus}{C$^+$}

\newcommand{\CII}{[C{\small II}]}
\newcommand{\COtwo}{CO(2--1)}

\newcommand{\COfive}{CO(5--4)}

\newcommand{\COseven}{CO(7--6)}

\newcommand{\HF}{HF J=1--0}
\newcommand{\HHO}{H$_{2}$O}
\newcommand{\HHOline}{H$_{2}$O\,(2$_{20}$-2$_{11})$}
\newcommand{\LHHO}{L$_\mathrm{H_{2}O\,(2_{20}-2_{11})}$}
\newcommand{\LHF}{L$_\mathrm{HF}$}
\newcommand{\LIR}{L$_\mathrm{IR}$}
\newcommand{\kms}{km\,s$^{-1}$}
\begin{document}

\authorrunning{Lehnert, Yang, Emonts et al.}

\title{Etching glass in the early Universe: Luminous HF and \HHO\ emission in a QSO-SMG pair at z=4.7\thanks{We are referring to the ability of hydrofluoric acid, a solution of water and hydrogen fluoride, to etch glass.}}

\titlerunning{Luminous \HF\ and \HHO\ emission from \src}

%\begin{CJK*}{UTF8}{gbsn}

\author{M.~D.~Lehnert\inst{1}\thanks{email: lehnert@iap.fr}
\and 
C.~Yang\inst{2}
\and
B.~H.~C.~Emonts\inst{3}
\and
A.~Omont\inst{1}
\and
E.~Falgarone\inst{4}
\and
P.~Cox\inst{1}
\and
P.~Guillard\inst{1}}

\institute{Sorbonne Universit\'e, CNRS UMR 7095, Institut
d'Astrophysique de Paris, 98bis bvd Arago, 75014 Paris, France
\and
European Southern Observatory, Ave. Alonso de C\'ordova 3107, Vitacura, Santiago, Chile
\and
National Radio Astronomy Observatory, 520 Edgemont Road, Charlottesville, VA 22903
\and
Laboratoire de Physique de l'ENS, \'Ecole Normale Sup\'erieure, Universit\'e PSL, CNRS, Sorbonne Universit\'e, Universit\'e Paris-Diderot, Paris, France}

\abstract{We present ALMA observations of hydrogen fluoride, \HF, water,
\HHOline, and the 1.2 THz rest-frame continuum emission from the z=4.7
system \src. System \src\ is a galaxy group consisting of a quasi-stellar
object (QSO), a sub-millimeter galaxy (SMG), and a pair of \La\ emitters. We detected HF in emission in the QSO and possibly in absorption in the SMG, while water
was detected in emission in both the QSO and the SMG. The QSO is the most luminous \HF\ emitter that has yet been found and 
has the same ratio of
HF emission-line to infrared luminosity, \LHF/\LIR,\ as a small sample of local active galactic
nuclei and the Orion Bar. 
This consistency covers about ten orders of magnitude in \LIR.
Based on the conclusions of a study of HF emission in the
Orion Bar and simple radiative transfer modeling, the HF emission in the QSO is  excited either
by collisions with electrons (and \Htwo) in molecular plasmas irradiated by
the AGN and intense star formation, or predominately by collisions with H$_2$, with a modest contribution from electrons, in a relatively high temperature ($\sim$120 K), dense ($\sim$10$^5$ cm$^{-3}$) medium. The high density of electrons necessary to collisionally excite the \HF\ line can be supplied in sufficient quantities by the estimated column density of \Cplus. Although HF should be an excellent tracer of molecular outflows,
we found no strong kinematic evidence for outflows in HF in either the QSO or the SMG.  From a putative absorption feature in HF observed against the continuum emission from the SMG, we conducted a bootstrap analysis to estimate an upper limit on the outflow rate, $\dot{\rm M}_{\rm outflow}\protect\la45$ \Msunyr. This result implies that the ratio of the molecular outflow rate to the star formation rate is $\dot{\rm M}_{\rm outflow}$/SFR$\protect\la$5\% for the SMG.
Both the QSO and the SMG are among the most
luminous \HHOline\ emitters currently known and are found to lie along the same relationship between
\LHHO/\LIR\ and \LIR\ as a large sample of local and high-redshift star-forming galaxies. The kinematics of the \HHOline\ line in the SMG is consistent with a rotating disk as found previously but the line profile appears
broader
than other molecular lines, with a full width at half maximum of $\sim$1020 \kms. The broadness of the line, which is similar to the width of a much lower resolution observation of CO(2-1), may suggest that either the gas on large scales ($\protect\ga$4 kpc) is significantly more disturbed and turbulent due either to interactions and mass exchange with the other members of the group, or to the dissipation of the energy of the intense star formation, or both.
Overall however, the lack of significant molecular outflows in either source may imply that much of the
energy from the intense star formation and active galactic nucleus in this pair is
being dissipated in their interstellar media.}


\maketitle

\section{Introduction}\label{sec:intro}

The evolution of galaxies is driven by the balance of energy and mass within
a baryonic gas cycle.  The factors that maintain this balance
are the rates of gas accretion from the cosmological web and mergers,
the angular momentum of the accreted gas, the star formation efficiency,
and outflows driven by starbursts and active galactic nuclei \citep[AGN;
e.g.,][]{lehnert15}.  Simulations of galaxies, especially high mass galaxies, suggest that a strong energy injection into the interstellar and circum-galactic media (ISM and CGM) is necessary to keep galaxies from growing overly massive, to ensure they have the correct age distribution of stellar populations, and to enable them to form the proper ratio of spiral and lenticulars as a function of redshift \citep[e.g.,][]{scannapieco04, dubois16, habouzit17, beckmann17}. Numerous studies have shown ample evidence for outflows from both galaxies with high star formation surface densities \citep[star formation rates per unit area; ][]{heckman90, lehnert96, beirao15} and AGN \citep[e.g.,][]{crenshaw12, cicone15, tombesi15}. However, outflows are only
one possible manifestation of starburst- or AGN-driven feedback.

Given the potential importance of feedback -- the self-regulating gas cycle through which galaxies and AGN limit their own growth -- it is important to understand what processes drive feedback and how the energy and momentum from young stellar populations and AGN is distributed and dissipated within the phases of the ISM and CGM \citep[see, e.g.,][and references therein]{guillard15, gray17, appleton18, buie18}. Understanding how the energy and momentum is distributed in bulk flows versus turbulence in gas, for example, provides insights into how feedback actually works in regulating galaxy and black hole growth.
To further our understanding of the
physics underlying the gas cycle in galaxies, especially outflows and
dissipation of energy generated by AGN and intense star formation, we used the Atacama Large sub/Millimeter Array (ALMA) to observe the galaxy
group \src\ at z=4.69 \citep{omont96, ohta96} in the \HF\ and para-\HHOline\ lines, whose rest frequencies are sufficiently close to be observed in a single tuning.

\src\ is a well-studied, unlensed group composed of a
quasi-stellar object (QSO), a sub-millimeter galaxy \citep[SMG; ][]{mcmahon94, smail97}, and two Ly$\alpha$
emitters \citep[LAEs; e.g.,][]{hu96}. Both the QSO and SMG are very luminous IR
emitters, $\sim$10$^{13}$\,L$_{\odot}$, and the system has been observed
in a wide range of molecular and atomic lines \citep[][and
references therein]{omont96, bedford99, salome12, carilli13, lu17a, lu18}.  
Based on an excess in the wing of the line profile of \CII, \citet{carilli13} estimated
an outflow rate in the atomic gas from the QSO of $\dot{{\rm
M}}_{\rm out}$$\sim$80\,\Msunyr\ and concluded that the gas
depletion time due to this outflow is $\sim$600 Myrs, a factor of $\ga$10 longer
than the gas consumption time due to star formation.  For the SMG, no outflows were inferred and the
\CII\ velocity field was interpreted as a rotating disk. The low
speed of the QSO outflow, a few 100 \kms, its low mass ejection
rate, and the lack of outflow in the SMG are very surprising given that
we expect QSOs to have fast winds, and starbursts that form stars
at greater than 1000\,\Msunyr\ to have vigorous outflows. If
neither the QSO and SMG are driving outflows, then perhaps the energy from
the intense star formation or from the luminous AGN is being rapidly
dissipated. Due to its nature, \HF\ is a good tracer of molecular outflows and \HHO\  traces
dissipation in molecular gas. When observed together, they enhance our understanding
of dissipation and the relative importance of outflows in \src\ in particular, and QSOs and SMGs generally.

Hydrogen fluoride\footnote{\citet{campbell79} and \citet{walker12} provide a discussion of the use and dangers of using HF (which can etch and dissolve glass) in an astronomical observatory to estimate stellar radial velocity variations.} has a large Einstein A coefficient and high critical
density, $\sim$10$^9$ cm$^{-3}$, implying that most of the HF gas
lies in its ground rotational state \citep{gerin16}. As a result, the 1-0 line is
generally observed in absorption in the Milky Way and other galaxies
\citep{neufeld97, neufeld05, neufeld10, rangwala11, monje11a, monje11b, monje14, kamenetzky12, P-S13, sonnentrucker15, P-B18}.  In
some sources, such as the Orion Bar and nearby galaxies hosting AGN, HF is observed in
emission \citep{vanderwerf10, vandertak12a, P-S13, lu17b, kavak19}. Little is currently known about HF emission or absorption in high redshift galaxies \citep[one is a detection in absorption and the other, only an upper limit; see][]{monje11b, lis11}.

Hydrogen fluoride is a robust molecule, representing the vast majority of the fluorine (F) in the
cool atomic and molecular phases of the ISM. F reacts exothermically with H$_2$, so it
rapidly forms HF \citep{neufeld09}. We note that the formation of HF through a reaction between F and H$_2$ has a moderate activation energy, E$_{\rm act}$/k$\sim$500 K, but the reaction rate is enhanced at low temperatures via quantum tunneling \citep{neufeld05}. HF is photo-dissociated only by photons
with $\lambda<$1120\AA, which means it is a robust molecule and shielded
even in relatively low column dusty neutral clouds, although it can be destroyed by reactions with C$^+$ \citep{neufeld09}. These characteristics
mean that HF is a sensitive probe of molecular gas columns over a wide range of
extinctions and densities, even in diffuse clouds \citep[A$_{V}<$0.5
magnitudes;][]{neufeld05} and because it traces the total \Htwo\ column, it is likely to be a sensitive probe of
even weak molecular outflows \citep[and inflows; e.g.,][]{monje14}. It is
only in dense, n$_{\rm H_2}$$\ga$10$^5$ cm$^{-3}$, cold, T$_{\rm gas}$$\la$20 K, molecular gas that HF may not be a good tracer of the total molecular
gas column due to adsorption onto grains \citep[``freeze-out''; e.g.,][]{neufeld05, vanderweil16}.
\HF\ is not a good tracer of turbulent dissipation \citep{godard14}.

Water is one of the main carriers of oxygen, after CO, in warm and cold
molecular gas.  Para-\HHOline\ is a relatively low excitation line, with an
upper energy level, E$_{\textrm{up}}$=195.9 K.  It has been detected in a significant
number of low- and high-redshift galaxies with a wide range of infrared luminosities
\citep[][and references therein]{yang13, yang16}. In contrast to CO,
\HHO\ is not a good tracer of photon-dominated regions (PDRs), because
it is easily photo-dissociated by UV radiation.  
Generally speaking, analyses of sources in which multiple transitions of \HHO\ have been observed, find that the excitation of \HHO\
is consistent with pumping by the infrared (IR) radiation field \citep{G-A12, G-A14}. The intense radiation field necessary to IR pump the water vapor emission may lead to an increase in the gas phase abundance of water by sublimation of
the icy mantles of dust grains. This increase in abundance may explain the high
luminosities, which  are generally beyond what is expected for PDR and IR pumping if much of the water was not in the gas phase \citep{G-A12, G-A14}. Other processes, such as exposure to intense, hard UV and X-ray radiation fields or the dissipation of mechanical energy, may also heat the high column density gas, melting the mantles of grains and increasing the rate of the formation of \HHO\ in the gas phase \citep[e.g.,][]{meijerink12}. In contrast to \HF, \HHO\ line emission is also a significant
source of energy loss and dissipation through slow molecular shocks \citep[v$_{\rm shock}$=20-40\,\kms\ and n$_{\rm H_2}$=10$^{3-5}$
cm$^{-3}$;][]{flower10, appleton13}. Consistent with this, although the emission is predominantly energized by IR pumping, water lines in nearby and
distant galaxies often show complex line profiles that generally have line
widths consistent with other molecular species \citep{G-A12, omont13}.

Our paper is organized as follows. In Sect.~\ref{sec:methods}, we present the observations, reduction, and analysis; in Sects.~\ref{sec:results} and \ref{sec:discussion}, we present the results and discuss their implications. In our analysis, we use a luminosity distance of 4.56$\times$10$^4$~Mpc and a physical scale of 6.8 kpc arcsec$^{-1}$.

%fig 1
\begin{figure*}[!ht]
\begin{center}
\includegraphics[width=0.95\linewidth]{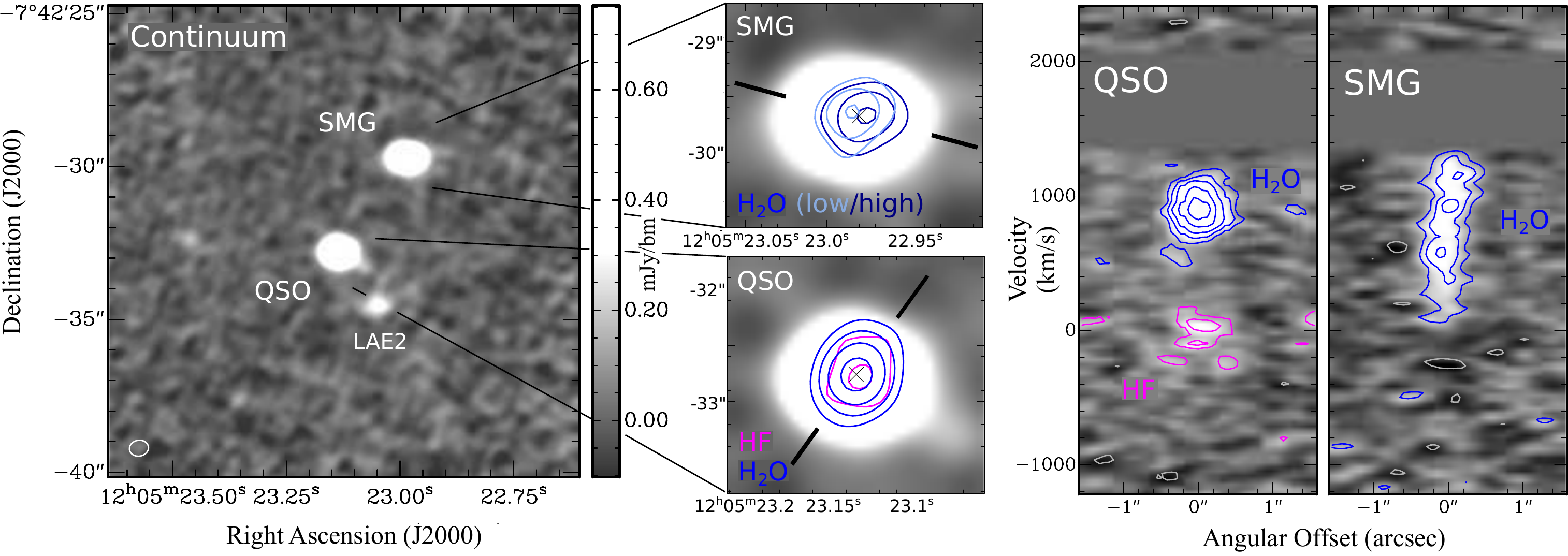}
\caption{\textit{left:} Total intensity image of the 1.2 THz rest-frame
continuum of both the QSO and SMG, along with \La\ emitter 2 in \src. The
half power beam size of the data, 0 \farcs56$\times$0 \farcs49 ($\sim$3.8\,kpc\,$\times$\,3.3\,kpc), is indicated in the lower left corner
of the panel. \textit{Middle:} Total intensity maps of the \HHOline\
emission in the SMG (top) and both the HF and \HHO\ emission in the
QSO (bottom), overlaid onto a gray-scale image of the continuum. For
the SMG, the \HHO\ at the lowest velocities (light blue; 20$-$915
\kms) and the \HHO\ at the highest velocities (dark blue; 915$-$1345
\kms) are kinematically resolved by $\sim$0.15$^{\prime\prime}$
($\sim$1\,kpc). For the QSO, both the \HHO\ (blue) and HF (magenta)
emission are unresolved. All velocities are with respect to the \HF\ line
assuming $z$\,=\,4.6948. The crosses mark the peak of
the continuum emission. Contour levels: 3, 6, 10, 15\,$\sigma$, with
$\sigma$ = 0.059 and 0.033 Jy\,beam$^{-1}$\,$\times$\,\kms\ for the low-
and high-velocity \HHO-emitting gas in the SMG, and $\sigma$ = 0.046 and
0.031 Jy\,beam$^{-1}$\,$\times$\,\kms\ for the \HHO\- and HF-emitting gas
in the QSO, respectively. \textit{Right:} Position-velocity plots of
\HHO\ (blue) and HF (magenta) emission in the QSO and SMG, taken along the
black lines in the middle plots. To increase the signal-to-noise ratio, we
binned every 6 channels of the data into a single channel and subsequently Hanning smoothed the binned data to a
velocity resolution of 66 km\,s$^{-1}$. Contour levels of the \HHO\ and HF
in both the QSO and SMG: $-$4, $-$2.3 (gray), 2.3, 4, 6, 9, 12 (blue/magenta)
$\sigma$, with $\sigma$\,=\,0.16 mJy\,beam$^{-1}$. The dark gray
region lies between the two spectral windows for which we have no data.
\label{fig:cont_line_images}}
\end{center}
\end{figure*}

\section{Observations and data reduction}\label{sec:methods}  %2

Our ALMA Cycle 3 observations in Band 6 were carried out on 2016\,March\,5
for 77\,minutes on-source integration time, with 36 antennas in the C36-3 configuration. The
four 1.875 GHz spectral windows were tuned to cover the frequency ranges 213-217.4\,GHz and
228-232.5\,GHz. The quasars, J1159-0940 and J1229+0203, were used to calibrate the complex gains and bandpass. The coverage of the visibility data in the u-v plane was well covered with baselines with lengths of
15m to 640m.  The source was observed at elevations of 58-77\degr\ and
the weather was stable with precipitable water vapor, PWV=1.6-2.2 mm.

We used the supplied calibration script
and Common Astronomy Software Applications (CASA; \citealt{mcmullin07}) to reduce the data. We imaged the phase calibrator and determined that the bandpass stability is accurate to $\sim$0.1$\%$. To image the line data, we ensured that no unwanted features were introduced across the band by first subtracting the continuum model from the visibilities. This reduced the continuum emission by an order of magnitude. We then subtracted the low-level residuals of the continuum by fitting a straight line to the line-free channels in the visibility data. To image both the continuum and line emission, we used natural weighting,
which resulted in a beam of 0 \farcs56$\times$0 \farcs49 with a
PA=$-$74$^\circ$ (Fig.~\ref{fig:cont_line_images}).

%fig 2
\begin{figure}[!ht]
\includegraphics[width=\linewidth]{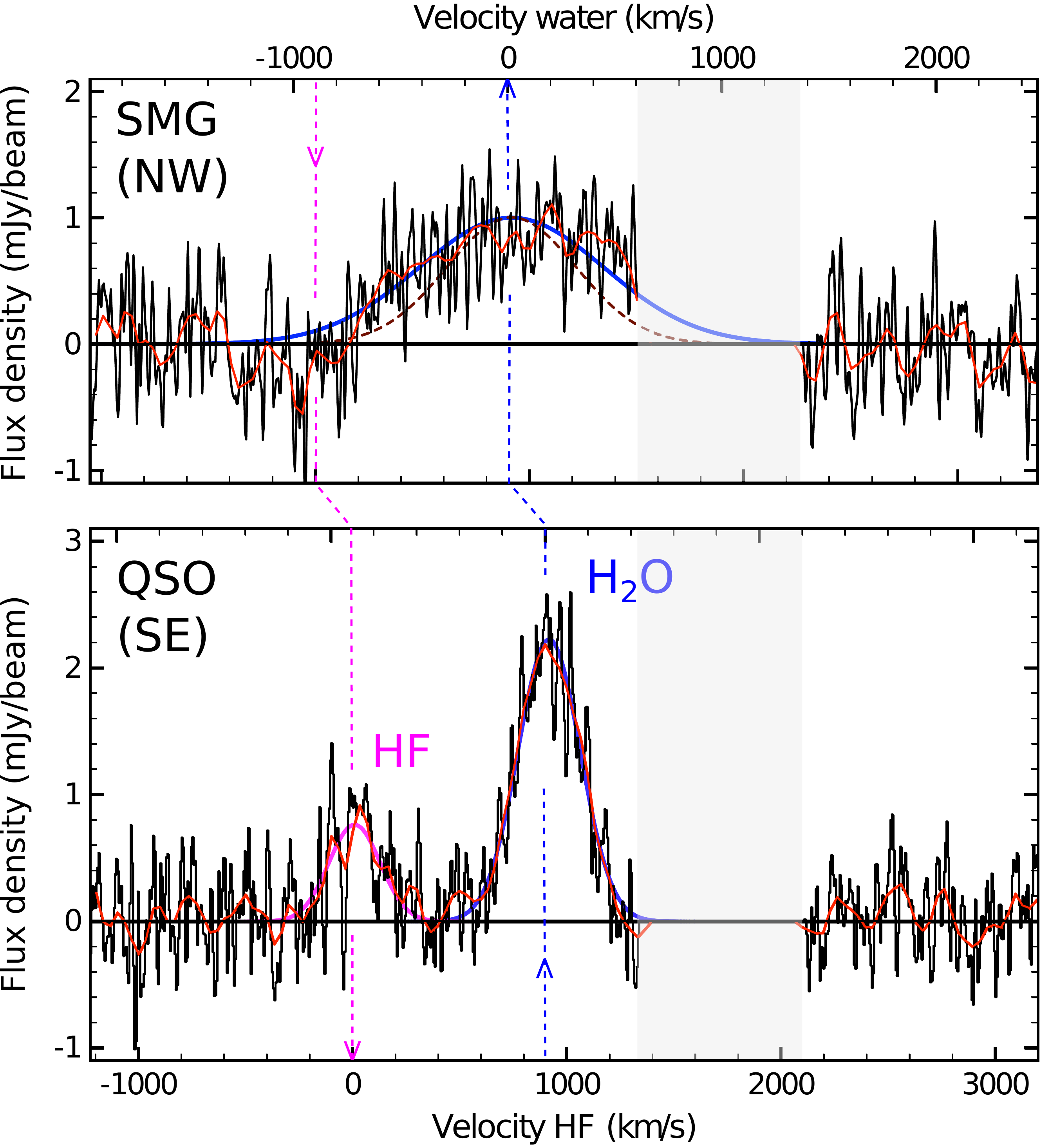}
\caption{Continuum-subtracted spectra of \HF\ and \HHOline\ in \src. The black line are
the data after being Hanning smoothed to a velocity resolution
of 11\,\kms. The red line shows the data binned into six
channels and subsequently Hanning smoothed to a velocity resolution
of 66\,\kms. \textit{Top:} spectrum of \HHO\ emission
from the SMG. Indicated with the magenta dashed line is the velocity of the expected \HF\ feature, which may (at best) be tentatively detected, relative to the velocity of \HHO. \textit{Bottom:} Spectrum of \HF\ and \HHOline\
emission from the QSO. The zero-velocity of the QSO is derived
from a Gaussian fit to the \HHO\ profile of the QSO, resulting in
$z$=4.6948$\pm$0.0001. The redshift of the SMG is assumed to be
$z$=4.6915 from \citet{carilli13}. The Gaussian functions represented with a solid line are fits to the HF (magenta) and H$_{2}$O (blue) emission, as summarized in Table~\ref{tab:results}. The Gaussian represented with the dotted brown line for the SMG is the fit to the [CII] line \citep{carilli13}. The horizontal axis on the bottom
(top) gives the corresponding velocity centered on the HF (\HHO)
line, as indicated with the magenta (blue) dashed line. The light gray region
indicates the gap in the spectra between the two spectral windows.
\label{fig:spectra}}
\end{figure}

%fig 3
\begin{figure}
\begin{center}
\includegraphics[width=\linewidth]{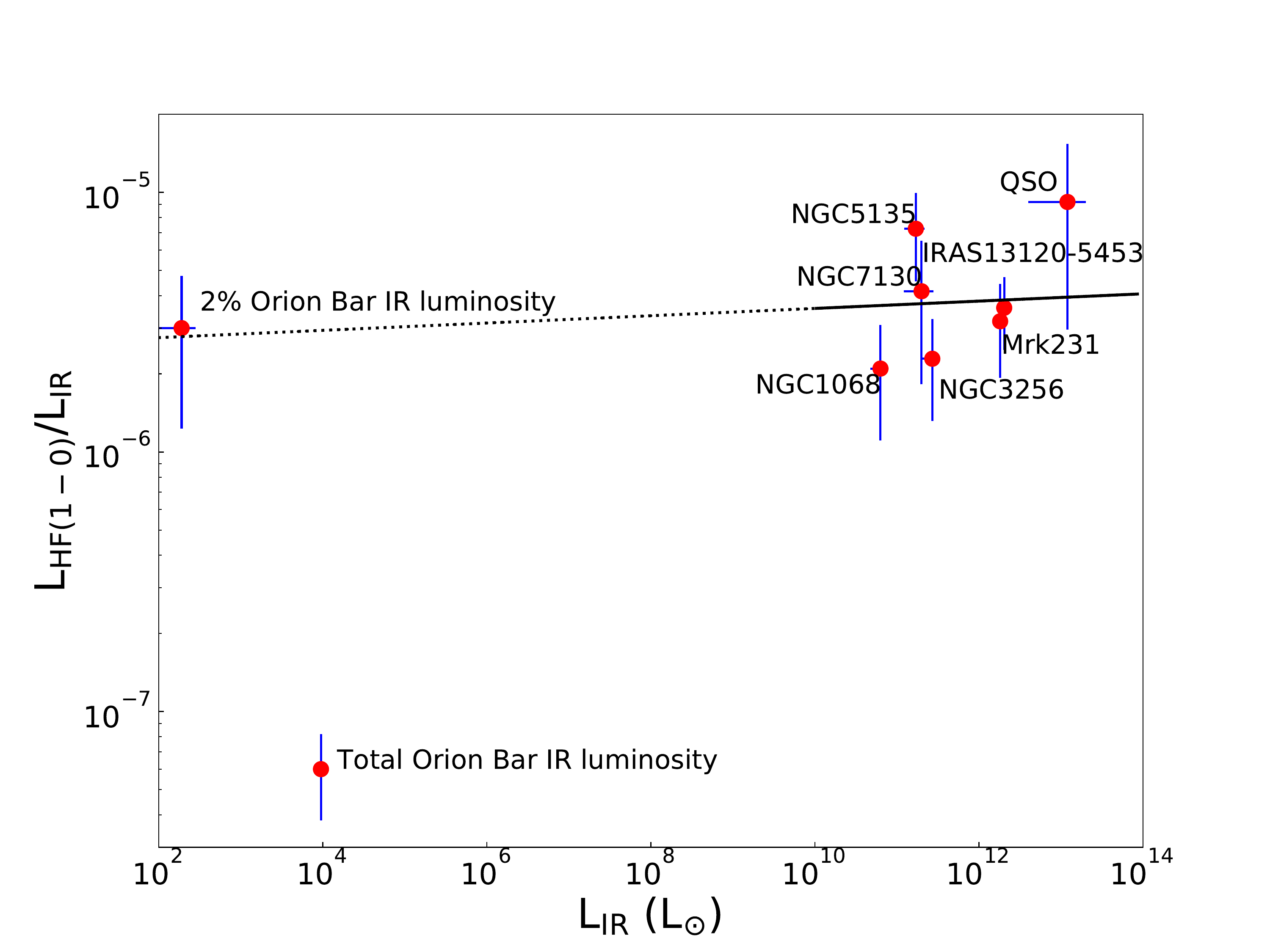}
\caption{Relationship between the infrared luminosity and the
ratio of the HF luminosity and infrared luminosity of \src, a sample
of local AGN \citep{yang13}, and the Orion Bar
\citep{salgado16,nagy17}. All sources are labeled. We show a
least-squares fit to the local AGN, QSO, and the Orion Bar (black line with the dotted line extension to the Orion Bar), which implies the ratio is constant.  We note that fitting the AGN and QSO only yields a similar fit but with a slightly steeper slope.
For comparison, we show the ratio \LHF/\LIR\ for the entire IR luminosity of the Orion
Bar and the luminosity if only 2\% of the total IR luminosity is within the Herschel beam used to
measure the HF flux (see text for details).
\label{fig:normLIRLHF}} 
\end{center}
\end{figure}

%table 1 Table~\ref{tab:specnumbers}
\setlength{\tabcolsep}{0.3em}
\begin{table}
\caption{Characteristics of the dust continuum and the H$_{2}$O and HF emission lines in \src.}
\label{tab:results}
\begin{centering}
\begin{tabular}{lcc}
\hline\hline\\[-2ex]      
 & QSO (SE) & SMG (NW) \\
\hline
\textbf{Continuum} & & \\
RA  & 12:05:23.134$\pm$0.001  & 12:05:22.980$\pm$0.001  \\
Dec & $-$07:42:32.76$\pm$0.01 & $-$07:42:29.680$\pm$0.01 \\
$S_{\textrm{216\,GHz}}$ (mJy) & 4.9$\pm$0.2 & 5.1$\pm$0.2 \\[1ex]
\textbf{\HHOline} & & \\[-2pt]
$z$& 4.6948$\pm$0.0001 & (4.6915)$^{\dagger}$ \\
$S_\textrm{peak}$ (mJy) & 2.2$\pm$0.1 & 1.0$\pm$0.1 \\
FWHM (\kms) & 340$\pm$15 & 1020$\pm$50 \\
$S\Delta$v (Jy\,beam$^{-1}$ \kms) & 0.80$\pm$0.03 & 1.1$\pm$0.2 \\[1pt]
\textbf{HF} & & \\
$S_{\rm peak}$ (mJy) & 1.0\,$\pm$\,0.1 & $-$ \\
FWHM (\kms) & 285$\pm$45  & $-$ \\
$S\Delta$v (Jy\,beam$^{-1}$ \kms) & 0.25$\pm$0.03 & $-$ \\
\hline
\end{tabular}
\end{centering}
\tablefoot{$^{\dagger}$ Because part of the H$_{2}$O line profile lies outside our frequency coverage, we constrained this to the \CII\ redshift \citep{carilli13}. The rest-frame frequencies of \HF\ and \HHOline\ are 1232.476 and 1228.789 GHz respectively.
}
\end{table}

\section{Results}\label{sec:results}

Both the QSO and SMG
are detected in the 1.2 THz rest-frame continuum, along with the \La\
emitter 2 (LAE2; Fig.~\ref{fig:cont_line_images}).
We detect \HF\ in emission
from the QSO and \HHOline\ in both the QSO and SMG (Fig.~\ref{fig:spectra}). 
The HF emission in the QSO is coincident
with the continuum emission and is spatially unresolved.
There is a weak absorption feature in the spectrum of the SMG over the velocity range
$\sim$0 to $-$400~\kms\ relative to the expected velocity of HF. We associate this weak putative absorption with \HF\  (Sect.~\ref{subsec:HFabsSMG}).
The \HHO\ emission is
unresolved in the QSO, but it is resolved in the SMG
(Fig.~\ref{fig:cont_line_images}). The position-velocity (PV) diagram
shows that the peak of the \HHO\ emission
shifts across $\sim$0.3$^{\prime\prime}$ or $\sim$2\,kpc, consistent with
the PV plot of \CII\ \citep{carilli13}.  Because of the large width
of the \HHO\ line in the SMG, the low frequency part of the profile
falls past the edge of our bandpass and some of the
emission is missing. Any potential \HHO\ emission at the redshift of LAE2 \citep{carilli13} would fall in the gap between the spectral windows.

Table~\ref{tab:results} summarizes the properties of the QSO and SMG. The continuum properties are derived by fitting a point source to
the unresolved 216\,GHz continuum of both the QSO and SMG. LAE2 has a 216 GHz
continuum flux density of 0.44$\pm$0.06 mJy beam$^{-1}$. The line
properties are derived by fitting a Gaussian function to the \HHOline\
and \HF\ lines. When we constrain the fit of the \HHOline\ in the SMG
to the redshift and width of the \CII\ line from \citet{carilli13}, we find
two regions of residual emission in both the blue
and red wing of the \HHO\ profile. Both of these residuals have approximately Gaussian shapes and a best fit to each have full widths at half maximum (FWHM) of $\sim$240$\pm$60 \kms.  The symmetry of the line and the residuals seems to rule out
significant emission from HF in the wing of the \HHOline\ line. Taken at face value, our results suggest that the water emission
is significantly broader than the other emission lines detected so far \citep[cf. 1020 \kms\ for \HHOline\ versus 720 \kms\ for the \CII, \COfive, and \COseven\ lines;][]{salome12,carilli13}.
We note that one observation, that of the \COtwo\ line with a beam of 2\,\farcs75$\times$1\,\farcs73,
has a \COtwo\ line FWHM comparable to the one we have estimated for the \HHOline\ line \citep{jones16}.
However, when we compare all of the line widths of \COtwo\ with restoring beam sizes of $\la$0 \farcs6 from \citet{jones16}, we find the  
weighted average FWHM=705~\kms. This implies that the most extended \COtwo\ emission has a larger line width.

In contrast, the line width of
the \HHOline\ in the QSO is, within the uncertainties, exactly the same
as for the other lines.  Neither the QSO or the SMG show clear evidence
for an outflow in the \HHO\ line, although there is a possible excess in the line
profile of the QSO seen $\sim$400 \kms\ blueward of the systemic \HHO\
redshift, consistent with the velocities of the weak \CII\ outflow
\citep{carilli13}.

\section{Discussion}\label{sec:discussion}  %3

\subsection{HF: Molecular gas irradiated by AGN and young stars}

The detection of HF emission in the QSO is unusual. In most of the
sources observed thus far,
the \HF\ line is observed as an absorption line with little or no emission \citep{vanderwerf10}. Only a handful of sources are known to have HF
purely in emission, without any obvious associated absorption. These sources are the Orion\,Bar \citep{vandertak12a, kavak19},
Mrk\,231 \citep{vanderwerf10}, and a few nearby galaxies \citep{P-S13, lu17b}.
To increase the number of galaxies observed with HF in emission,
we extracted the HF emission line fluxes from local galaxies observed with Herschel in the
sample of \citet{yang13}, which all host AGN.  We find that
the galaxies that host AGN and the QSO in \src\ have an approximately constant ratio
of \LHF/\LIR\ irrespective of \LIR\ (Fig.~\ref{fig:normLIRLHF}). We note, however, that the exact slope of the relation only considering the local AGN and QSO is dependent on the far-infrared luminosity used. Other estimates of the infrared luminosity of the QSO in the literature would lower \LIR/\LHF\ \citep[cf.][]{salome12, wagg14, lu17a}.

% add about how the L_HF were derived (Solomon and Downes) and extracted from Herschel data

The Orion Bar is the only known galactic source or sight-line where
HF appears purely in emission \citep[although we note that the absorption of
HF and the nearby \HHOline\ could mask any emission in other galactic and extra-galactic sources; see e.g.,][]{monje14, neufeld10, sonnentrucker15}. If we
compare \LHF/\LIR\ and \LIR\ for the Orion Bar \citep{salgado16, nagy17}, we find
that its \LHF/\LIR\ lies over an order of magnitude below. However,
the beam over which the HF flux is extracted from the Orion\,Bar
region only subtends $\sim$2\% of the projected area of the 250$\mu$m flux in the Orion\,Bar
\citep[cf.][]{salgado16}. If we scale the IR luminosity of the
Orion Bar by this factor, we find that it has approximately the same
ratio as the QSO in \src\ and the other AGN in our sample. However, the beam used to measure the
\HF\ flux does not subtend a simple geometric projection of the flux at 250$\mu$m and thus our
estimate of 2\% may be too low by a factor of approximately two.

In their analysis of the HF emission in the Orion Bar, \citet{vandertak12a}
found that collisions with electrons was the likely excitation mechanism,
requiring n$_{e-}$$\sim$10~cm$^{-3}$. This high electron density in the
molecular gas, and the fact that at the time the other source known
to have HF emission was Mrk\,231, led them to speculate that other AGN
would have HF in emission. 
We confirm this speculation and suggest that the constant ratio of \LHF/\LIR\ in such sources that we observe
simply reflects the high intensity of energetic photons that the AGN and intense star formation
provide, raising the electron density and perhaps excitation temperature in the
molecular gas to levels necessary to excite HF emission \citep{vandertak12a, kavak19}.

To understand why the QSO in \src\ has a similar \LHF/\LIR\ as the Orion Bar, we estimated the 
UV photon densities.  Considering the combined
contributions from the AGN and star formation for the QSO and the
star formation for the SMG, we estimate a non-ionizing radiation intensity
of $>$300 G$_0$ and $\sim$520 G$_0$\footnote{To make this estimate, we used the star formation rates
\citep[from \LIR,\ which was estimated over the wavelength range of 20-1000~$\mu$m;][]{salome12}, the UV continuum slope, the flux density at 1550~\AA\
for continuous star formation at an age of 5 Myrs estimated using Starburst99
\citep{leitherer99}, and the radius of the continuum emission from our
study for the QSO (an upper-limit) and the 44 GHz size for the SMG
\citep{jones16}. For the non-ionizing radiation from the QSO itself,
we used the estimate of the 1450\AA\ continuum flux density and the
continuum slope from \citet{carniani13}. In all estimates of the non-ionizing continuum, we integrated the scaled UV
continua from 6--13.6 eV. The estimate for the 
QSO is lower than that for the SMG due to the larger size used to estimate the intensities
in the QSO compared to the SMG.}. The non-ionizing radiation
intensity is $\sim$10$^4$ G$_0$ in the Orion Bar \citep{Hh95}, about an
order of magnitude higher than the SMG and the lower limit for the QSO.
Similarly,
we find that for the QSO (SMG), the density of ionizing photons
is $\ga$900 cm$^{-3}$ ($\sim$200 cm$^{-3}$), assuming all the ionizing photons have an energy
of 13.6 eV. Using the results from
\citet{odell17}, the ionizing photon density in the Orion\,Bar is $\sim$60
cm$^{-3}$. In the QSO, the AGN is about a factor of
four more luminous in its ionizing radiation than that due to its star formation, and
likely has a much harder radiation field.
The globally large ionizing photon intensity and the likely high
G$_0$ in the QSO implies that there is sufficient photon
intensity in the diffuse molecular gas to maintain a high electron density. This agreement may be
fortuitous given the crudeness of our estimate.  We certainly cannot rule out the (likely) contribution
from X-rays and cosmic rays in ionizing and heating the \HF\ emitting regions. Both X-rays and cosmic rays may be necessary to penetrate deeply enough to provide a sufficient volume of \HF\ emitting gas to explain the strength of the \HF\ line.

To investigate the similarity of the QSO \HF\ emission to that of the Orion\,Bar, we used the code RADEX \citep{vandertak07}\footnote{https://personal.sron.nl/$\sim$vdtak/radex/index.shtml.} to constrain the column density and excitation of the \HF\ line. We tried a variety of models to explain the brightness temperature of the HF emission, including ones used previously to model the CO emission in \src\ \citep{salome12}.  To make an estimate of the peak temperature of the emission, we assumed that the source size was that of the beam of the ALMA observations of \src. If we used the high angular resolution that was obtained of the dust continuum, $\sim$0 \farcs3 \citep{salome12}, it would increase the brightness temperature by about a factor of three. The estimated temperature of the dust continuum is about 43 K \citep{salome12}. The model parameters that give the appropriate brightness temperature are the ones close to those used in \citet{vandertak12a}, namely, the column density, N(HF)=10$^{15}$ cm$^{-2}$, the number density of molecular hydrogen, n$_{\rm H_2}$=10$^{5}$ cm$^{-3}$, the excitation temperature, T$_{\rm gas}$=43 K, background cosmic microwave background temperature, T$_{\rm bg}$=15.5 K, and an electron number density, n$_{\rm e^-}$=10 cm$^{-3}$. In these calculations, we assume a turbulent velocity dispersion of 5~\kms\ as observed in the Orion Bar \citep{vandertak12a, nagy17}. If we assume a higher velocity dispersion, the column density necessary to explain the strength of the \HF\ emission would increase proportionally. If we use the dust continuum size, 0 \farcs3, then the brightness temperature is best explained with a higher excitation temperature, T$_{\rm gas}$=100 K. The high excitation temperature is the same as that used to model the \HF\ emission for a region in the Orion Bar \citep{vandertak12a}. Consistent with this possible higher excitation temperature for a more compact emission in the QSO, recently \cite{kavak19} found that the \HF\ emission in the Orion Bar is consistent with a higher excitation temperature, T$_{\rm gas}$$\sim$120 K, and a molecular density, 10$^5$ cm$^{-3}$. With this density and temperature, the excitation of \HF\ is dominated by collisions with \Htwo\ with only a modest contribution to the excitation from electrons of about 15\%. We also find that this could explain the HF emission in the \src\ QSO but only if the HF emitting region is compact. Future high resolution observations can test whether this is the case.

Thus it appears that perhaps the regions of molecular gas in the QSO host galaxy are similar to that in the Orion Bar, but on a much larger scale. The total line width of the \HF\ line, since it is very similar to the other relatively large number of molecular and atomic lines observed in \src, is due to a large number of individual clouds orbiting within the gravitational potential of the QSO host galaxy.

There are other mechanisms that could potentially excite HF emission
that are not considered in the radiative transfer modeling, including near-IR pumping and residual energy from the formation of HF molecules \citep[``chemical pumping'';][]{vandertak12b, godard13}. However, the radiation field intensities of 1000-10$^5$ times the interstellar radiation field would only increase the population of the J=1 of HF by a small amount \citep{godard13}. The impact of chemical pumping is more
difficult to estimate given the limited constraints we have on the radiation field impinging on and the density distribution of the HF-bearing molecular gas \citep{vandertak12b}. Assuming an equilibrium between the formation and destruction of HF \citep[see][for details]{vandertak12b}, we estimated the required column of molecular gas necessary to explain the strength of the \HF\ line emission in the \src\ QSO.  We find that the total molecular column density, N$_{\rm H_2}$, must be $\ga$10$^{24}$ cm$^{-2}$ to explain the total column density of HF in the J=1 rotational level.  The necessary column density of molecular gas is well above that estimated in \citet{salome12} based on radiative transfer modeling of the strengths of several CO lines (and what we estimate in the following). Thus we can rule out a significant contribution by chemical pumping to the excitation of HF.

Electron densities are an important factor in connecting the Orion Bar and our observations of the QSO in \src. This is because collisions with electrons is a potentially important mechanism for exciting \HF\ emission and whose importance depends on the precise excitation temperature of the gas \citep[cf.][]{vandertak12a, kavak19}. The electron fraction is 10$^{-4}$ from the RADEX modeling. To check if that is a plausible value, we used RADEX to model the brightness temperature of the \CII$\lambda$\,158$\mu$m line \citep{carilli13}. We find that the brightness temperature of the \CII\ can be explained by a similar model to that used to explain the strength of the HF line, but with a lower excitation temperature, T$_{\rm ex}$=55 K, and a column density, N(\Cplus)=4$\times$10$^{18}$\,cm$^{-2}$. Except for T$_{\rm ex}$, which is about a factor of two smaller (and about that used to model the CO emission in \citealt{salome12}, namely a dust temperature of 43\,K), these parameters are almost identical to those found by \citet{vandertak12a} to model the Orion Bar.  We note that since the \CII\ line is optically thick, increasing the column does not increase the brightness temperature.  The only way to do that is to increase the excitation temperature. If the \Cplus\ to H$_2$ density ratio is about 10$^{-4}$ (consistent with the solar abundance of C), then, as expected in models of PDRs, C would be able to supply the necessary density of electrons in the low extinction regions of the PDRs in the QSO. At higher columns, cosmic rays (and to a much less extent, turbulent dissipation) may also increase the electron densities \citep{meijerink05,meijerink11,godard13}.

HF is thought to be an excellent tracer of the total \Htwo\ gas column density since it probes molecular gas even at relatively low levels of extinction \citep{neufeld05, neufeld09}. However, to estimate the total molecular column density from HF we need to know the relative abundance of HF in the molecular gas.  The reaction of F with molecular hydrogen is exothermic and thus needs no energy source to facilitate its formation \citep[][and references therein]{gerin16}. Thus, we expect that almost all of the fluorine is in the form of HF in the molecular gas, as observed in other sources. If we make that assumption and further assume that the abundance of fluorine is solar \citep[relative abundance, F/H=3.6$\times$10$^{-8}$;][]{asplund09}, then the total molecular column is
$\sim$10$^{22}$ cm$^{-2}$.  Interestingly, assuming the same for the carbon abundance \citep[C/H=2.69$\times$10$^{-4}$;][]{asplund09} results in a similar total molecular column density \citep[see the CO column density estimates given in][]{salome12}.  

While this agreement could be fortuitous, it is consistent with the gas being metal enriched, but moreover suggests that fluorine was produced rapidly.  Although uncertain, models of fluorine production and observations suggest that the majority of the F is produced in massive stars and a significant fraction may also be produced in low to intermediate mass stars in their asymptotic branch phase of evolution \citep{cunha04, cunha08, abia10, abia11, abia15, prantzos18}. The production of F peaks in AGB stars with $\sim$2 \Msun\ and is produced in stars with masses over the range \textasciitilde1-3 \Msun\ \citep{lugaro08}. The evolutionary time for stars to become AGB stars over this mass range is $\sim$0.6-2 Gyr. The age of the universe is only $\sim$2 Gyr at the redshift of \src.  How the fluorine abundance was enhanced given the evolutionary timescale of the stars that produce a significant fraction of the F is an interesting questions. That the production of fluorine relies on the abundance of oxygen and nitrogen is also of note. More observations are needed to constrain the abundance of F in galaxies in the early universe in order to understand what our observations imply about fluorine nucleosynthesis.

\subsection{Possible detection of HF in absorption in the spectrum of \src\ SMG}\label{subsec:HFabsSMG}

\HF\ is frequently observed in absorption
\citep{monje11a, monje11b, monje14, sonnentrucker15} and since it probes the total
column of molecular gas over a wide range of extinctions and densities, it is an
excellent probe of outflowing molecular gas \citep[][]{monje14}. Therefore, if either the QSO or SMG were driving outflows, we would expect to see strong broad HF absorption.
The spectrum of the SMG shows possible absorption over the velocity range of about 0 to $-$400~\kms\ relative to the systemic velocity for the \HF\ line (Fig.~\ref{fig:spectra}). The significance of this feature is not high, but if real, its characteristics are what we would expect to observe if the SMG was driving an outflow.  To assess the significance of this feature, its column density, and mass outflow rate, we analyzed it via a bootstrap method assuming it is HF in absorption.
We did 1000 realizations of the data by binning the unsmoothed data by a factor of ten to a velocity sampling of 60~\kms. We estimated the standard deviation of the channels within each bin and used this as the distribution of the uncertainty in the flux density of each binned channel. We converted all of the flux densities in each channel to optical depths, multiplied by the channel width, and summed the resulting depths over the velocity range 0 to $-$400~\kms. For these estimates, we assumed that all of the HF was in the ground state, which given the nature of HF is reasonable. In this case, the optical depth is simply $\tau$=-ln(F$_{\rm line}$/F$_{\rm continuum}$). Using the relation from \citet{neufeld10}, we estimated the integrated column density for each realization.

In addition, we calculated the outflow rates based on each estimated column density. To estimate the outflow rates, we used the relation for an expanding wind with an opening angle of $\pi$ sr \citep[e.g.,][]{heckman90, lehnert96}, $\dot{\rm M}_{\rm wind}$=$\pi$\,m$_{\rm H_2}$\,N(H$_2$)\,V$_{\rm wind}$\,R$_{\rm wind}$, where m$_{\rm H_2}$ and N(H$_2$) are the mass and column density of molecular hydrogen respectively, V$_{\rm wind}$ is the terminal velocity of the wind (which we assume to be 400~\kms), and R$_{\rm wind}$ is the injection radius of the wind (which we assume to be the half beam size of the ALMA data). This relation assumes the outflow is steady and launched at radius, R$_{\rm wind}$, with terminal velocity, V$_{\rm wind}$. To convert between the column density of HF and H$_2$, we assumed an abundance of HF as inferred in the Orion Bar, 3.5$\times$10$^{-8}$ \citep{vandertak12a}. 
From this analysis, we find that 95\% of the estimated values for the column density of HF and the mass outflow rates are 
N(HF)$\la$4$\times$10$^{13}$ cm$^{-2}$ and $\dot{\rm M}_{\rm wind}\la$45 \Msunyr.  The star formation rate of the QSO is \textasciitilde10$^{3}$ \Msunyr\ \citep{salome12}. Thus the wind efficiency of only the star formation is $\dot{\rm M}_{\rm wind}$/SFR$\la$5\%. This is much less than typically found in starburst and AGN host galaxies \citep{fluetsch19}.

There are several caveats that we need to recognize in estimating the upper limit of the outflow rates.  We assume that the fluorine abundance is solar (i.e., similar to that inferred for the Orion Bar).  If it is much less, than the upper limit in the molecular outflow rate would be proportionally higher. We also do not know the launch radius of any potential outflow in the \src\ SMG and have simply assumed the ALMA beam size.  If the source is much more compact, the limits will be proportionally lower. To observe the \HF\ line in absorption, it must be viewed against the thermal dust continuum and therefore the dust continuum size is appropriate for the launch radius. 

There are
also perhaps two more mundane explanations for not observing strong absorption in the SMG of \src. The first
is that the HF molecules could be in equilibrium with the thermal radiation field. In this case,
there would be no emission or absorption. If the molecular gas has a low density, sufficiently low such that collisional excitation of HF is not important, being in equilibrium with the radiation field implies that the
rate of absorption of the thermal continuum is equal to spontaneous and stimulated emission. Assuming a detailed
balance and a radiation field temperature of 43 K, we find that the radiation intensity necessary at the
frequency of \HF\ is over two orders of magnitude greater than that observed. An alternative and perhaps  plausible explanation
could be that the SMG is driving a wind but we are observing a disk edge-on. In an edge-on disk, there is no bright continuum emission against which the outflowing gas is superposed and therefore no absorption is observed. Any emission is likely too faint to be detected as the inner regions would be swamped by the thermal emission from the disk. A similar effect is observed in nearby galaxies in the Na~D absorption lines in the optical whereby as a disk galaxy becomes more edge-on to the line of sight, the Na D absorption is dominated by gas in the ambient ISM and not by gas from an outflow \citep{heckman00}. This last possibility can be tested with higher resolution observations to determine the morphology and axial ratio of the dust continuum, as well as the kinematics of the SMG, to determine whether it is
an approximately edge-on rotating disk.

\subsection{\HHO: IR-pumping and energy dissipation}

\begin{figure}
\begin{center}
\includegraphics[width=\linewidth]{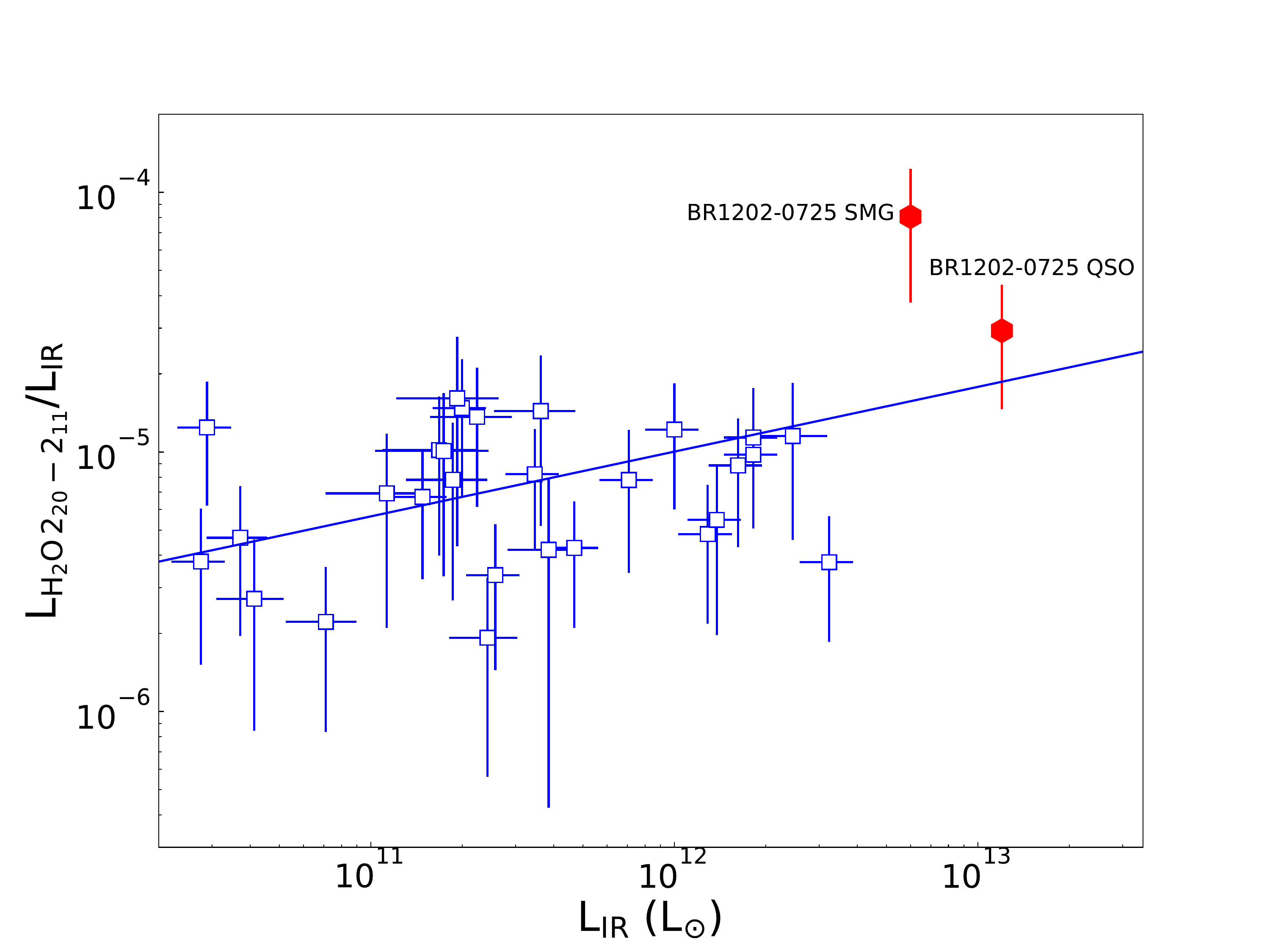}
\caption{Relationship between the infrared luminosity, \LIR, and the
ratio of the \LHHO\ and \LIR\ of \src\ (red hexagons) and for a sample of local galaxies \citep[blue hollow squares][]{yang13}.
We show a
least-squares fit to all galaxies (blue line, slope of 0.19$\pm$0.02). To make the estimates for the QSO and SMG, we used the infrared luminosities from \citet{salome12}.  Other estimates from
the literature are a few times higher \citep{wagg14, lu18} and would yield values that have greater consistency with the best fit shown in the figure.
\label{fig:normLIRLH2O}} 
\end{center}
\end{figure}

We find that the QSO and SMG lie above a relation
between the \LHHO/\LIR\ and \LIR\ with a slope of 0.19$\pm$0.02
\citep[Fig.~\ref{fig:normLIRLH2O}; see][and also \citealt{lui17}]{yang13}.
\citet{yang13} found a best fitting slope for the relationship between \LHHO\ and \LIR\ that is consistent with what we find, which is not surprising given we are using the same data and only including two additional points. 
The high \LHHO/\LIR\ of both the QSO and SMG of \src\ indicate
that they are among the most, if not the most, luminous emitters of \HHOline\
currently known. The relationship between \LHHO\ and \LIR\ is consistent with
IR pumping of the water lines
\citep[][]{G-A14, yang13, yang16}. The water abundance is estimated to be an order of magnitude higher for galaxies that lie along the relationship between \LIR\ and water luminosity
\citep[$\sim$10$^{-6}$;][]{G-A10} than the water abundance in
the Orion Bar \citep[$<$few 10$^{-7}$;][]{habart10,nagy17}. The \HHOline\ line is not detected in the Orion Bar \citep[e.g.,][]{nagy17}.
The small slope of the \LHHO/\LIR--\LIR\ relation may suggest that the gas-phase abundance of water also increases as the IR luminosity increases.

%fig 4

However, the water line in the SMG
appears broader than the lines of the other dense gas tracers
\citep[about 40\% wider;][]{salome12}. \citet{jones16}, analyzing observations of \src\ with the Jansky Very Large Array, found that the \COtwo\ line width at beam sizes $\la$0\,\farcs6 was roughly constant at $\sim$700\,\kms \citep[see also][]{salome12}. When they analyzed the data with a restoring beam size of 2\,\farcs6$\times$1\,\farcs7, the \COtwo\ FWHM increased to $\sim$1050\,\kms\ or similar to what we estimate for the \HHOline\ line. Since our beam is about 0.5 arcsec, we would expect the line to have a FWHM of $\sim$700\,\kms, especially given that the \HHOline\ emission is not spatially resolved.  Thus it could also be that the line widths of different tracers become larger at larger scales. The water-emitting gas could also be more sensitive to the specific process or processes that are causing the line width to increase with increasing size of the restoring beam size \citep[see][for details]{jones16}.

There are several plausible explanations for this effect with decreasing spatial resolution.
The detection of broader lines with decreasing resolution may indicate that the rotation curve is rising on scales larger than about 0\,\farcs5 ($\sim$3.4 kpc at the redshift of \src).
Unlike \HF, water emission also traces
dissipation in slow molecular shocks \citep{flower10}. In the SMG, there could
be 
mechanical energy dissipation causing emission in addition to (the dominant excitation mechanism) IR-pumping, consistent with 
the broad line observed in the water line and its high luminosity.  
The SMG, despite having about half the IR luminosity of the QSO
\citep{jones16}, is 40\% brighter in \HHOline\ \citep[see][for analyses suggesting that star-forming galaxies hosting AGN have less luminous water emission]{omont13,yang13}.  \src\ is an interacting system with evidence of a bridge of gas connecting the QSO and SMG \citep{carilli13}.  Such an interaction and potentially significant gas mass exchange between the two galaxies would excite gas and induce complex kinematics, perhaps preferentially on large scales \citep[see discussion in][]{emonts15}.
Moreover, if the kinematics and luminosity of the water lines are related to the dissipation
of energy in the interstellar media of the SMG, it may explain why we observe only relatively weak outflows; much of the mechanical energy generated by the
intense star formation is not contributing to driving outflows, but is being dissipated
in the dense molecular gas. All of these processes -- infrared pumping, turbulent dissipation of the mechanical energy from the young stars, the transfer of mass from or to the QSO, or even outflows -- may be contributing to exciting the broad water emission in the SMG. Unfortunately, the source of this energy to support the large line width on large scales, or the underlying cause of the high \HHOline\ luminosity are not constrained by our data. Observations of additional water lines and other tracers of dissipation in dense molecular gas such as CH$^+$, SH$^+$, and rotational levels of \Htwo\ \citep{godard14} at a variety of spatial resolutions
will be necessary to substantiate any of these various hypotheses.

\begin{acknowledgements} 
MDL wishes to thank Nikos Prantzos for interesting discussions on the nucleosynthetic origin of fluorine.
This paper makes use of the following ALMA data: ADS/JAO.ALMA \#2015.1.01489.S.
ALMA is a partnership of ESO (representing its member states), NSF (USA), and NINS (Japan),
together with NRC (Canada), MOST and ASIAA (Taiwan), and KASI (Republic of Korea), in cooperation with the Republic of Chile. The Joint ALMA Observatory is operated by ESO, AUI/NRAO, and NAOJ.
The National Radio Astronomy Observatory is a facility of the NSF operated under cooperative agreement by AUI. CY acknowledges support from an ESO Fellowship. EF acknowledges support from the European Research Council in the form of the Advanced Grant MIST (FP7/2017-2022, No. 742719). We thank the anonymous referee for their constructive comments that helped us to improve this manuscript.
\end{acknowledgements}

%\bibstyle{aa}
\bibliographystyle{aa}
\bibliography{BRI1202}
%\end{CJK*}
\end{document}